\documentclass[review,preprint,12pt,fleqn]{elsarticle}%

\usepackage[marginparwidth=20mm]{geometry}
\usepackage{amsmath,amssymb,amsthm, mathtools}

\DeclareMathOperator*{\argmin}{arg min}
\newcommand\tsr[1]{\boldsymbol{#1}}
\newcommand\pderiv[3][]{\frac{\partial^{#1} #2}{\partial #3^{#1}}}

\usepackage{microtype}
\usepackage{lmodern}
\usepackage{anyfontsize}
\usepackage{natbib}
\usepackage{siunitx}
\usepackage[backgroundcolor=gray!50, linecolor=white]{todonotes}

\todostyle{sintef}{bordercolor=blue, color=blue!30, author={Sintef}, inline}
\newtheorem{pbm}{Problem}
\newtheorem{rmk}{Remark}
\usepackage{array,booktabs}

\usepackage{tikz}
\usetikzlibrary{arrows,calc,patterns}
\usetikzlibrary{external}
\tikzexternaldisable
\usepackage{shellesc}
\usepackage{pgfplots}
\usepackage{pgfplotstable}
\usepgfplotslibrary{colormaps}
\pgfplotsset{compat=1.15}
\usepgfplotslibrary{groupplots}
\definecolor{viri01}{HTML}{440154}
\definecolor{viri01}{HTML}{440154}
\definecolor{viri02}{HTML}{482173}
\definecolor{viri03}{HTML}{433E85}
\definecolor{viri04}{HTML}{38598C}
\definecolor{viri05}{HTML}{2D708E}
\definecolor{viri06}{HTML}{25858E}
\definecolor{viri07}{HTML}{1E9B8A}
\definecolor{viri08}{HTML}{2BB07F}
\definecolor{viri09}{HTML}{51C56A}
\definecolor{viri10}{HTML}{85D54A}
\definecolor{viri11}{HTML}{C2DF23}
\definecolor{viri12}{HTML}{FDE725}
\definecolor{sintefblue}{HTML}{003C65}
\definecolor{sintefblue}{RGB/cmyk}{0,60,101/1,.57,0,.4} 
\definecolor{sintefgrey}{RGB/cmyk}{235,235,230/0,0,0,.1} 
\colorlet{sintefgray}{sintefgrey}
\definecolorset{RGB/cmyk}{sintef}{}{lightgreen, 205,250,225/.23, 0,.20, 0;%
                                    green,       20,185,120/.73, 0,.67, 0;%
                                    darkgreen,    0, 70, 40/.93,.43,.92,.52} 
\definecolorset{RGB/cmyk}{sintef}{}{yellow, 200,155,20/20, 36,98, 8;%
                                    red,    190, 60,55/19, 86,77, 8;%
                                    lilla,  120,  0,80/48,100,27,31} 
\definecolorset{HTML}{sintef}{}{cyan,      22A7E5;%
                                magenta,   EC008C;%
                                lightgrey, D8D0C7} 
\colorlet{sinteflightgray}{sinteflightgrey}

\journal{Open review}
\begin{document}

\title{%
 Reformulating plastic instabilities within a damage-like variational framework
}

\author[sintef]{Baptiste Reyne\corref{corauth}}
\cortext[corauth]{\texttt{baptiste.reyne@sintef.no}}

\affiliation[sintef]{%
 organization={SINTEF},
 country={Norway},
}

\begin{abstract}
 Yield point phenomena in metals, anticracks in soils and snows, or planewise collapses in lattice metamaterials, 
 are different manifestations of plastic instabilities characterized by stress softening and strain localization.
 Several independent constitutive descriptions exist and suffer from similar challenges: experimental difficulties to link local events to global measurements, and computational lack of convergence, spurious localization and mesh sensitivity.
 This work proposes a minimal field-agnostic constitutive framework where plastic softening is represented by a damage-like internal variable which explicitly separates stable hardening behavior from unstable softening.
 The resulting formulation admits a variational structure and can leverage regularization tools originally developed by the ductile fracture community.
 The framework aims at emphasizing similarities within plastic instability phenomena and establishing a bridge with continuum damage mechanics, to open a path towards regularized variational formulations of localized plastic flows.
\end{abstract}



\def\myLF{Lip-field}
\def\myPF{Phase-field}



\maketitle

\def\myd{\par\rule[2pt]{10pt}{2pt}~}
\clearpage
\section{Introduction}


A plastic deformation process is termed \emph{stable} when strains develop homogeneously under homogeneous loads.
Plastic instabilities arise when the same uniform loads cause inhomogeneous deformation.
\begin{itemize}
 \item In metal where plasticity happens a near-constant volume, they can manifest as localization bands: this is shear banding, L\"uders bands or the Portevin--Le-Chatelier effect \cite{rowlands2023}.
 \item In soils and snows that can plastically deform with volume change, they take the form of compaction bands or \emph{anticracks} \cite{barraclough2016,gaume2018}.
 \item In architected materials such as 3D printed metallic lattice structures, local buckling can appear macroscopically as planewise collapse (or \emph{layer-by-layer}) or yielding \cite{tancognedejean2016}.
\end{itemize}
Even in the absence of plasticity, elastic instabilities can arise as a consequence of local buckling causing planewise collapses in visco-elastic architected elastomers \cite{bertoldi2008,meijer2026}.
The common trait of these phenomena is, upon loading, the occurrence of a peak resistance followed not by fracture, but by softening and localized deformation.
On global measurements this typically translates into a plateau that in many contexts is characteristic of a phase change%
\footnote{In our example cases, phase change can be an alloy before and after strain aging, or a granular or architected medium before and after compaction.}.
Considering the similarity of these softening behaviors, the modeling and simulation of plastic instabilities could welcome a common description that emancipates from small-scale phenomena.

%


%
For plastic instabilities, softening behaviors are traditionally modeled not as separable phenomena, but as negative hardening.
It can be done directly as a single hardening curve showing a momentary negative slope~\cite{marais2012}, or implicitly through negative strain rate sensitivity (NSRS), that is loss of resistance with increasing strain rate as used by \citeauthor{gaume2018}\cite{gaume2018} for snow or \citeauthor{manach2014}\cite{manach2014} for an aluminium--magnesium alloy.
Other approaches involve dedicated internal variables and micro-structure inspired evolution models.
For example the PLC effect is often modelled using the Kubin--Estrin--McCormick (KEMC)~\cite{kubin1985,mccormick1988} framework where softening is modelled by a sudden loss of resistance driven by an internal variable representing the dislocations aging time.
Although expressed in units of time, this variable's evolution does control softening in a straightforward manner as its sudden drops lead to corresponding momentary losses of yield stress in the model.
In general these macroscopic descriptions do not allow to separate hardening (stable) from softening (unstable) behaviors since they both respond to a positive strain rate.


The continuum damage and ductile fracture community addresses a similar problem using a different track: there we introduce a dedicated independent internal variable to measure softening and model its consequences as a loss of strength or stiffness.
This softening variable $d\in[0,1]$ represents the smooth transition between a healthy material and a fully damaged one.
An important aspect is that models are often built following some aspects of generalized standard materials \cite{halphen1975}: the local constitutive equations all derive from a single scalar potential.
This gives the model a variational structure that enables a number of robust and efficient computational solving methods \cite{bourdin2008}.
A first important framework is variational fracture or \myPF{} \cite{miehe2010} that regularize the problem with  gradient penalization interpreted as an energy contribution.
The more recent \myLF{} approach yields comparable outcomes without the use of gradients of internal variables, but rather by imposing a Lipschitz-continuity constraint to the softening variable\cite{moes2021,gopalsamy2023}.


The goal of the present work is to improve plastic instability models by adapting their structure to ductile fracture tools and algorithms through a minimal proof of concept.
The model must be able to capture softening events followed by a potential re-hardening.
In its description, it must distinguish hardening phenomena from softening ones using dedicated independent variables. 
Finally, in order to maximize the possibilities of solvers and regularization approaches, it must admit a variational structure or be compatible with gradient approaches.


To do so, the present paper proposes a model--solver couple following the \myLF{} framework~\cite{moes2021,chevaugeon2022} adapted to softening plasticity.
In Section~\ref{sec:model} a minimal model is formulated locally, brought to one dimension and regularized.
While the compatibility with various regularization methods is shown, a \myLF{}--compatible solver is followed, illustrated then analyzed in Section~\ref{sec:discussion}.
The broader ambition of this work is to promote the elaboration of a common computational toolbox that exploits the similarities of all softening mechanics.

%
%

\section{Model--solver formulation}
\label{sec:model}

\subsection{Minimal recurrent softening model}

The constitutive framework is built in accordance with the following requirements.
\begin{enumerate}
 \item The behavior represented by the model reproduces stress softening and strain localization: while the material globally hardens, the growth of plastic strains implies localized softening associated with a stress drop. 
  This softening is bounded by a plastic increment magnitude that gives a limit to the stress drop.
 \item The model is based on a state potential from which derive all variables so that the constitutive problem admits a variational formulation.
  The variables evolution can be written as equality or inequality constraints to the corresponding minimization.
 \item The model variables are split into two groups respectively associated to the material hardening and the material softening.
  The state potential is not strictly convex, but is independently convex with respect to these two groups.
\end{enumerate}

The following model satisfies the above conditions with the minimum possible complexity.
We consider a particle of an elasto-plastic solid undergoing scalar strain loads.
The corresponding first variable is
\begin{itemize}
 \item the strain $\epsilon\in\mathbb R$,
\end{itemize}
The internal variables used in this model are
\begin{itemize}
 \item the plastic strain $\epsilon_p\in\mathbb R$ such that the difference $\epsilon-\epsilon_p$ quantifies reversible kinematics;
 \item the accumulated plastic strain $p$ such that $\mathrm dp = |\mathrm d\epsilon_p|$;
 \item and the softening variable $d\in[0,1]$.
\end{itemize}
We introduce the bar operator
\begin{align}
 \left.
  \begin{array}{rll} 
  \bar p:\mathbb R^+&\longrightarrow& [0,1]\\
  p&\longmapsto&\dfrac{\min{(p,\delta)}}{\delta}
 \end{array}
 \right]
\end{align}
that represents the evolution of a localization event as $\bar p$ goes from 0 to 100\% when $p$ makes a step from  0 to $\delta$, the latter representing the strain developed during the localization event.
This is a simple and general model to help describing a softening event starting at yield, using the following state potential.

The local constitutive problem is described by the potential
\begin{equation}
 \label{eqn:pizero}
\pi_0 
= 
\underbracket{ \frac12 E(\epsilon-\epsilon_p)^2}_{\text{elastic } \pi_e}
+ \underbracket{ \sigma_y\left(p+\frac12kp^2\right)}_{\text{hardening } \pi_h}
+  \underbracket{ \delta\left(1-d\right)\bar p\sigma_a + \delta\frac12d^2\sigma_a }_{\text{softening } \pi_s}
\end{equation}
where parameters include the Young modulus $E$, the yield stress $\sigma_y$, the normalized linear hardening rate $k$ and the aging stress $\sigma_a$. 
The latter corresponds to the strength initially added and then lost during the localization event (no change happens for $\sigma_a=0$, a plateau is obtained for $\sigma_a=\sigma_y k \delta$, and softening is observed with higher values).
The potential yields the state equations
\begin{align}
 \sigma =\pderiv{\pi}{\varepsilon}&=E(\epsilon-\epsilon_p), \label{eqn:state_sig}\\
 R     =\pderiv{\pi}{p}          &=  \sigma_y(1+kp) + \pderiv{\bar p}{p}(1-d)\delta\sigma_a\label{eqn:state_r},\\
 \mu   =\pderiv{\pi}{d}          &= \delta(-\bar p + d) \sigma_a\label{eqn:state_mu}.
\end{align}
The minimization procedure would inherently cancel out these quantities since they are derivatives.

We introduce a time discretization by considering known all solutions at time increment $t_n$, and seeking solutions at time $t_{n+1}=t_{n}+\Delta\! t$.
For the sake of brevity and without loss of clarity, we drop the $(n+1)$ index and only conserve the $(n)$ index to refer to the previous increment.

We impose the following constraints associated with the variables' evolution.
\begin{align}
 \mathcal E_0 = \left\{ 
  (\epsilon, \epsilon_p, p, d)\in \mathbb R^{4}:
  \left[
  \begin{array}[h]{rl}
   |\sigma(\epsilon, \epsilon_p)| - R(p,d) &\leq0,\\
   p - p_n &=  |\epsilon_{p}-\epsilon_{p,n}|
  \end{array}
 \right.
 \right\}
\end{align}
where $R$ follows \eqref{eqn:state_r}.
They correspond to the yield criterion (inequality constraint) and the accumulated plastic strain positive rate condition (equality constraint).
\begin{rmk}
 The softening variable evolution is bound to the accumulated plastic strain $p$ through $\mu=0$ from equation \eqref{eqn:state_mu} which is inherently handled by the minimization of $\pi$.
\end{rmk}

For a single material point, the kinematics--driven constitutive problem consists in imposing the strain and can be summarized as follows. 
\begin{pbm}[local, non-regularized]
 \label{pbm:local}
 Given imposed strains $\epsilon_{n+1}\equiv\epsilon$, find
 \begin{align} 
  (\epsilon_p, p, d) = \arg\min_{(\bullet,\epsilon_p^\prime, p^\prime, d^\prime)\in\mathcal E_0}
  \pi(\epsilon,\epsilon_p^\prime, p^\prime, d^\prime).
 \end{align}
\end{pbm}
This constrained problem brings stable solutions as shown below, although its Hessian (Appendix \ref{sec:app}) is already only conditionally positive definite: the potential is non-convex by construction, which is a desired source of localization.

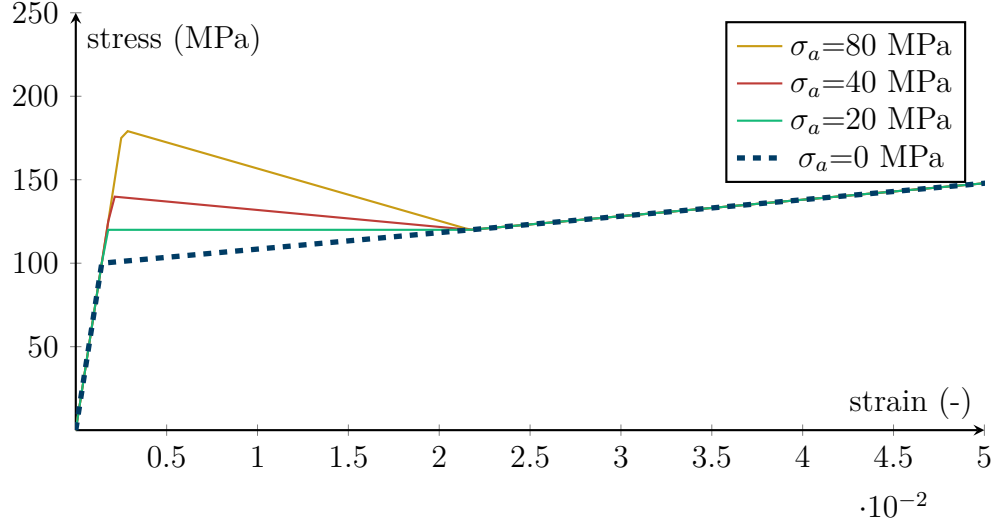
\begin{figure}[ht]
 \centering
 \pgfplotscreateplotcyclelist{MyCycleList}{%
   {darkgray, mark = none, thick},
   {Green, mark = none, thick},
   {MidnightBlue, mark = none, thick},
   {Mahogany, mark = none, thick},
   {RedOrange, mark = none, thick},
   {RoyalPurple, mark = none, thick},
 }
 \begin{tikzpicture}
  \begin{axis}[
   axis lines=middle,
   ymin=0, ymax=250,
   xmin=0, xmax=.05,
   width=.9\linewidth,
   height=.47\linewidth,
   thick,
   xlabel=strain (-),
   ylabel=stress (MPa),
  ]
  \addplot[sintefyellow]table{dat/drop_80.0.dat}; \addlegendentryexpanded{$\sigma_a$=80~MPa}
  \addplot[sintefred]table{dat/drop_40.0.dat}; \addlegendentryexpanded{$\sigma_a$=40~MPa}
  \addplot[sintefgreen]table{dat/drop_20.0.dat}; \addlegendentryexpanded{$\sigma_a$=20~MPa}
  \addplot[line width=2pt, sintefblue, dashed]table{dat/drop_0.0.dat}; \addlegendentryexpanded{$\sigma_a$=0~MPa}
  \end{axis}
 \end{tikzpicture}
 \caption{
  Simulation results of the local model (problem~\ref{pbm:local}) with $\sigma_y$=100~MPa, $k$=10, $\delta$=0.02, $E$=70~GPa, and $\sigma_a$ set from 0 to 80~MPa.
 }
 \label{fig:zerod}
\end{figure}

Figure \ref{fig:zerod} presents the work-hardening behavior numerically computed for the present model.
It can be observed that the behavior requirements are reproduced in a minimal and efficient way: a parametrized strain step accompanied by a controlled stress drop.


\subsection{Extension to one dimension}
We now consider a bar under tension--compression.
The bar $\Omega$ of constant section is discretized into $(N-1)$ elements of constant length connected at nodes $(x_i)_{1\leq i\leq N}$.
This spatial discretization is formally described by the approximation of the displacement field
\begin{align}
 U(x) \approx u(x) = \mathcal N_i(x) u_i\quad\text{where the nodal values}\quad u_i=U(x_i).
\end{align}
The shape functions $\mathcal N_i(x)$ yield a linear interpolation hence the piecewise constant strain approximation
\begin{align}
 \varepsilon(\tilde u) = \pderiv{u}{x}
 \approx
 \varepsilon(u(x)) = \pderiv{\mathcal N_i}{x} u_i
\end{align}
that yields, at each element centroid (integration point)
\begin{align}
 \varepsilon_k(u_{k},u_{k+1}) = \varepsilon(\tilde u(x_k+\Delta x/2)) = \frac{u_{k+1}-u_{k}}{\Delta x}
\end{align}
with $k$ the element index and $(u_k,u_{k+1})$ the connected nodes' displacements.

\begin{rmk}
 From thereon we agree to use fields' names as a shorthand for the corresponding collection of discretized variables.
 For example, $u\equiv(u_i)_{1\leq i\leq N}$ at nodes and $p\equiv(p_k)_{1\leq k\leq N-1}$ at integration points.
\end{rmk}

The potential at each integration point reads
\begin{equation}
 \pi_{k}(u_{k},u_{k+1}, \epsilon_{p, k}, p_{k}, d_{k}) = \pi_0(\varepsilon_k(u_{k},u_{k+1}), \epsilon_{p, k}, p_{k}, d_{k})
\end{equation}
and its integral over the bar:
\begin{align}
 \int_\Omega \pi_0 \mathrm d\! V
 \approx A\Delta x\sum_{1\leq k\leq N_\text{el}} \pi_{k}
 \equiv \Pi(u, \epsilon_p, p, d)
\end{align}
with $A\Delta x$ a constant element volume.
Finally we control the problem by splitting the nodal displacement into imposed and unknown:
\begin{align}
 \mathcal U = \left\{ 
  u\in \mathbb R^{N}: 
  u_1 = 0,\quad u_{N}=u_\text{imp}(t)
  \right\}
\end{align}
and rewrite the evolution constraints to take the discretization into account
\begin{align}
 \mathcal E_1 = \left\{ 
  (u,\epsilon_p, p, d)\in \mathbb R^{4N-3}:
  \left[
  \begin{array}[h]{rl}
   |\sigma(\varepsilon(u), \epsilon_p)| - R(p,d) &\leq0,\\
   p - p_n &=  |\epsilon_{p}-\epsilon_{p,n}|
  \end{array}
 \right.
 \right\}.
\end{align}

\begin{pbm}[global--discretized, non-regularized]
 Given solutions at increment $n$ and $u_\text{imp}(t_{n+1})$, find:
 \begin{align}
 (u,\epsilon_p, p, d) = \argmin_{
  \substack{
   u^\prime\in \mathcal U\\
   (u^\prime,\epsilon_p^\prime, p^\prime, d^\prime)\in\mathcal E_1\\
  }
 }
 \Pi(u^\prime, \epsilon_{p}^\prime, p^\prime, d^\prime)
 \end{align}
\end{pbm}

This problem is ill-posed since the softening variable induces spurious localization. 
In general there are several admissible solutions:
\begin{itemize}
 \item a succession of localized mesh sensitive strain bursts accompanied by an elastic unload over the rest of the body;
 \item or a homogeneous evolution of internal variables with successive load drops: satisfying solutions that cannot be reached with standard minimization.
\end{itemize}
In the following section we deploy the \myLF{} approach \cite{moes2021,chevaugeon2022} to both regularize the problem and efficiently find solutions.

\clearpage

\subsection{One-dimensional solver with spatial regularization}
This approach proposed by \cite{moes2021} is inspired by the Lipschitz continuity definition, which requires that there exists a constant K such that the following bound is satisfied by a function $d$ for any couple of points $(\tsr x, \tsr y)$.
\begin{align*}
 |d(\tsr x)-d(\tsr y)|\leq K\mathrm{dist}(\tsr x, \tsr y)
\end{align*}
where $\mathrm{dist}$ satisfies the properties of a common metric, or distance function: positive for different points, nil for identical points, symmetric and verifying the triangular inequality.
From the modelling standpoints, this is an efficient way to impose some degree of regularity without the need to compute a gradient, or even to impose differentiability.

The \myLF{} approach states that an admissible softening variable field $d$ belongs to
\begin{align}
 \mathcal L = 
 \left\{
  d\in\mathcal L^\infty:
  |d(x_i)-d(x_j)|\leq \frac{|x-y|}{\ell}\quad\forall x,y\in\Omega
 \right\}
\end{align}
where $\mathcal L^\infty$ is the space of essentially bounded functions as used in the reference study~\cite{chevaugeon2022} and where $\ell$ represents an internal length.
The discretized Lip-constraints in the bar example uses the collection of integration points $(g_k)_{1\leq k\leq N-1}$ and the corresponding values $d_k$.
It reads
\begin{align}
 \left\lbrace
 \begin{array}{cccc}
  \frac{-|g_1 - g_{2}    |}{\ell}\leq d_1 - d_{2}    \leq  \frac{|g_2 - g_{1}    | }{\ell}\\
  \ldots \\                                       
  \frac{-|g_{k-1} - g_{k}|}{\ell}\leq d_{k-1} - d_{k}\leq \frac{|g_{k} - g_{k-1}| }{\ell}\\
  \ldots \\                                       
  \frac{-|g_{N-2} - g_{N-1}|}{\ell}\leq d_{N-2} - d_{N-1}\leq \frac{|g_{N-2} - g_{N-1}| }{\ell}
 \end{array} 
 \right.
\end{align}
which is implemented using the following linear operators.
\begin{align}
 \begin{bmatrix}
     -|g_2-g_1|/\ell \\
     -|g_3-g_2|/\ell \\
     \ldots         \\
     \ldots         \\
     -|g_{N-1}-g_{N-2}|/\ell \\
     -\infty \\
 \end{bmatrix}
 \leq
 \begin{bmatrix}
     1 &  -1 &        &        &   &   \\
       &   1 &  -1    &        &   &   \\
       &     & \ldots &        &   &   \\
       &     &        & \ldots &   &   \\
       &     &        &        & 1 & -1\\
       &     &        &        &   &  1 
 \end{bmatrix}
 \begin{bmatrix}
     d_1 \\
     d_2 \\
     \ldots         \\
     \ldots         \\
     d_{N-2} \\
     d_{N-1}
 \end{bmatrix}
 \leq
 \begin{bmatrix}
     |g_2-g_1|/\ell \\
     |g_3-g_2|/\ell \\
     \ldots         \\
     \ldots         \\
     |g_{N-1}-g_{N-2}|/\ell \\
     \infty \\
    \end{bmatrix}.
\end{align}
This is a linear constraint that can be handled efficiently by optimization algorithms.

\begin{pbm}[global--discretized, regularized]
 \label{pbm:reg}
 Given solutions at increment $n$ and $u_\text{imp}(t_{n+1})$, find:
 \begin{align}
 (u,\epsilon_p, p, d) = \argmin_{
  \substack{
   u^\prime\in \mathcal U\\
   (u^\prime,\epsilon_p^\prime, p^\prime, d^\prime)\in\mathcal E_1\\
   d^\prime\in\mathcal L\\
  }
 }
 \Pi(u^\prime, \epsilon_{p}^\prime, p^\prime, d^\prime)
 \end{align}
\end{pbm}

The problem is now formulated in order to prevent spurious localization, but the minimization remains compromised because the potential remains non-convex.
To tackle this we split the non convex minimization into two convex steps: hardening and softening.
This is done using a staggered scheme as promoted in \cite{chevaugeon2022}.
\begin{enumerate}
 \item Given solutions at increment $n$ and $u_\text{imp}(t_{n+1})$, initialize $d\leftarrow d_n$.
 \item (hardening iteration) Fix the softening variable and find
 \begin{align}
 (u,\epsilon_p, p) = \argmin_{
  \substack{
   u^\prime\in \mathcal U\\
   (u^\prime,\epsilon_p^\prime, p^\prime, \bullet)\in\mathcal E_1\\
  }
 }
 \Pi(u^\prime, \epsilon_{p}^\prime, p^\prime, d).
 \end{align}
 \item (softening iteration) Fix the previous solutions to find
 \begin{align}
 d = \argmin_{
  \substack{
   (\bullet, \bullet, \bullet, d^\prime)\in\mathcal E_1\\
   d^\prime\in\mathcal L\\
  }
 }
 \Pi(u, \epsilon_{p}, p, d^\prime)
 \end{align}
 which represents the \myLF{} projection
\item repeat until stagnation of the converged solutions.
\end{enumerate}

\section{Results and discussion}
\label{sec:discussion}

\paragraph{One-dimensional behavior}
The solver for problem~\ref{pbm:reg} is implemented in \cite{gitlabplasmat} and used to simulate a one-dimensional tensile test discretized in 15 nodes.
The parameters are fixed to $\sigma_y=100$~MPa, $\sigma_a=40$~MPa, $k=10$, $E=70$~GPa, $\delta=0.02$ and only the internal length left to vary for 1/1000 to 2.
A defect is added in the center element in the form of an initial plastic strain.

\begin{figure}
 \newcommand\MyInGr[2]{
  \begin{tikzpicture}
   \node (pic) at (0,0){
    \includegraphics[width=65mm,trim={5mm 0mm 15mm 15mmm},clip]{images/K10_SIGa40_del2_15bars_ell#1_WIRE.png}
   };
   \path (pic.north) node[below, fill=white, fill opacity=.5, text opacity=1]{#2};
  \end{tikzpicture}
 }
 \centering
 \MyInGr{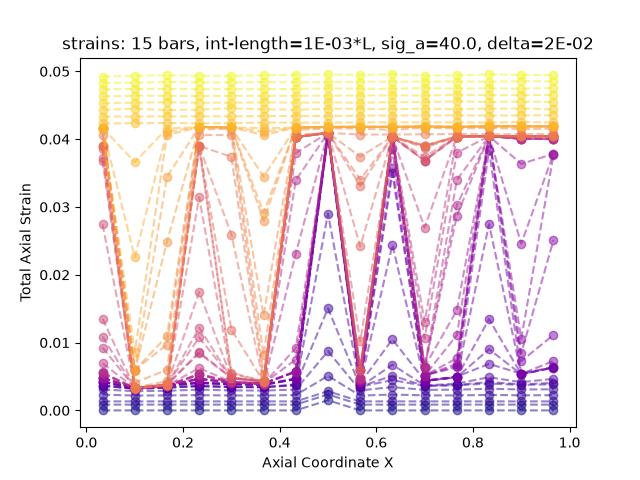}{(a) $\ell/L$ = 0.1\%}
 \MyInGr{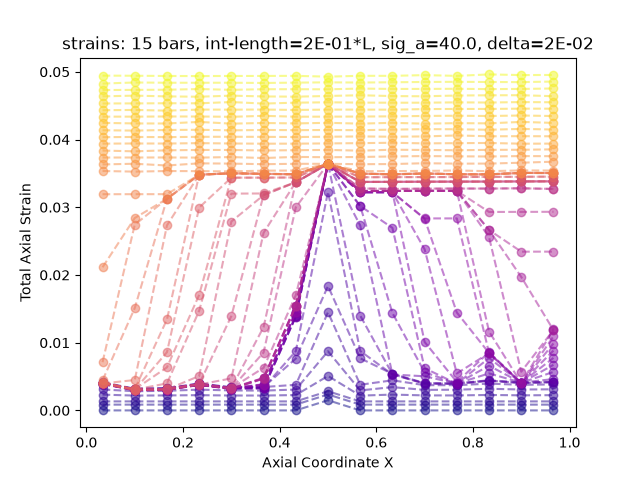}{(b) $\ell/L$ = 25\%}
 \MyInGr{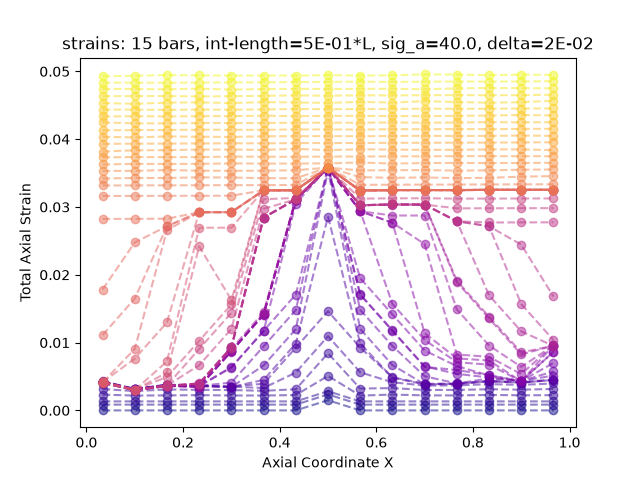}{(c) $\ell/L$ = 50\%}
 \MyInGr{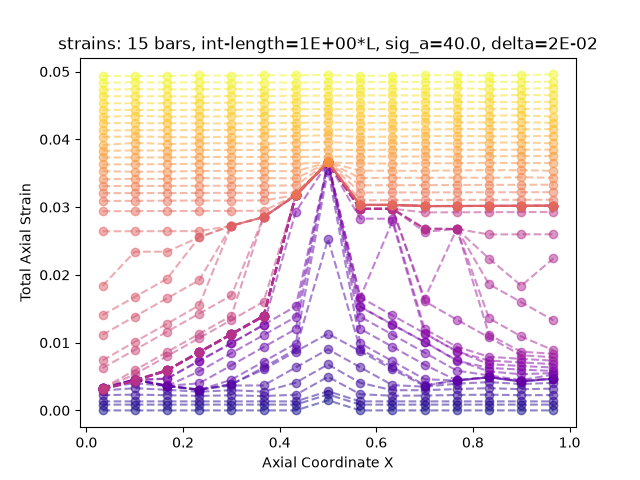}{(d) $\ell/L$ = 100\%}
\caption{
 Influence of the regularization length over the results of a 15-node bar under tension.
 The x--axis is that of the bar and the results show the evolution of the total strain at integration points.
}
 \label{fig:oned}
\end{figure}
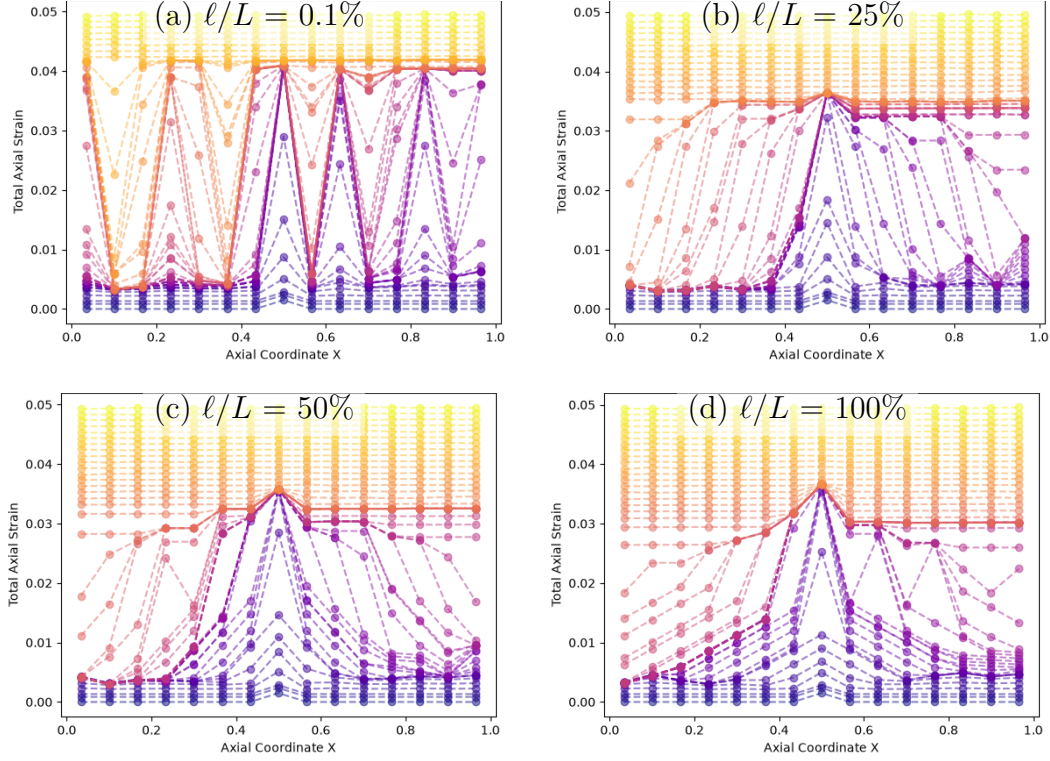

Figure~\ref{fig:oned} presents four simulation cases of the 15-node bar under imposed elongation from 0 to $L/20$.
The graphs feature strain distributions where colors (dark to light) correspond to the simulation increment.
As it increases, the solution becomes less and less localized and the strains progress more and more homogeneously.
In a general context, this corresponds to the width of localization bands that increases as the softening variable's Lipschitz--continuity is constrained.
The first localization is always observed in the center of the bar where the defect is introduced, then propagates towards the right before propagating to the left in every regularized computation.
The lip constraint does not appear explicitly on the strain distribution but is strongly imposed on the softening variable, hence the plastic part of the strains as long as $d\leq1$.

Figure~\ref{fig:oned}-a features a regularization length that is significantly smaller than the size of a finite element: it carries no influence over the solutions that localize spuriously without any close neighbourhood dependencies.
In Figure~\ref{fig:oned}-b and further, the internal length becomes larger than a finite element and effectively regularizes the solutions.

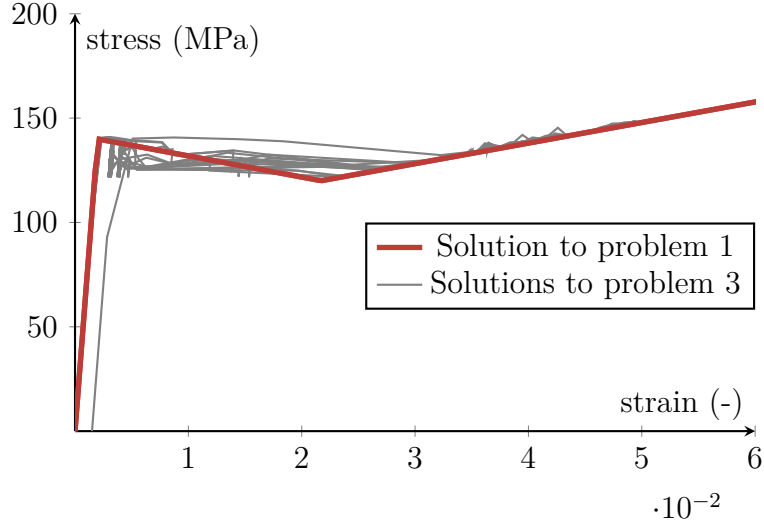
\begin{figure}[h]
 \centering
 \begin{tikzpicture}
  \begin{axis}[
   axis lines=middle,
   ymin=0, ymax=200,
   xmin=0, xmax=.06,
   width=.7\linewidth,
   height=.47\linewidth,
   thick,
   xlabel=strain (-),
   ylabel=stress (MPa),
   legend style={at={(1.,0.5)}, anchor=north east,}
  ]
  \addplot[line width=2pt, sintefred]table{dat/drop_40.0.dat};
  \addplot[solid, gray, ] table[]{dat/hardening_K10_SIGa40_del2_15bars_ell25_elt0.dat};
  \addplot[solid, gray, ] table[]{dat/hardening_K10_SIGa40_del2_15bars_ell25_elt1.dat};
  \addplot[solid, gray, ] table[]{dat/hardening_K10_SIGa40_del2_15bars_ell25_elt2.dat};
  \addplot[solid, gray, ] table[]{dat/hardening_K10_SIGa40_del2_15bars_ell25_elt3.dat};
  \addplot[solid, gray, ] table[]{dat/hardening_K10_SIGa40_del2_15bars_ell25_elt4.dat};
  \addplot[solid, gray, ] table[]{dat/hardening_K10_SIGa40_del2_15bars_ell25_elt5.dat};
  \addplot[solid, gray, ] table[]{dat/hardening_K10_SIGa40_del2_15bars_ell25_elt6.dat};
  \addplot[solid, gray, ] table[]{dat/hardening_K10_SIGa40_del2_15bars_ell25_elt7.dat};
  \addplot[solid, gray, ] table[]{dat/hardening_K10_SIGa40_del2_15bars_ell25_elt8.dat};
  \addplot[solid, gray, ] table[]{dat/hardening_K10_SIGa40_del2_15bars_ell25_elt9.dat};
  \addplot[solid, gray, ] table[]{dat/hardening_K10_SIGa40_del2_15bars_ell25_elt10.dat};
  \addplot[solid, gray, ] table[]{dat/hardening_K10_SIGa40_del2_15bars_ell25_elt11.dat};
  \addplot[solid, gray, ] table[]{dat/hardening_K10_SIGa40_del2_15bars_ell25_elt12.dat};
  \addplot[solid, gray, ] table[]{dat/hardening_K10_SIGa40_del2_15bars_ell25_elt13.dat};
  \addplot[solid, gray, ] table[]{dat/hardening_K10_SIGa40_del2_15bars_ell25_elt14.dat};
  \addplot[line width=2pt, sintefred]table{dat/drop_40.0.dat};
  \addlegendentryexpanded{Solution to problem 1}
  \addlegendentryexpanded{Solutions to problem 3}
  \end{axis}
  \end{tikzpicture}
  \caption{
   Axial tensile stress vs.\ total axial strain corresponding to Figure~\ref{fig:oned}-b ($\ell/L$ = 25\%) plotted for each element of the bar.
   The reference line shows the corresponding ``zero--dimensional solution''  (Figure~\ref{fig:zerod} with $\sigma_a$=40~MPa).
  }
 \label{fig:localhard}
\end{figure}


\paragraph{Algorithmic aspects}
The current implementation uses \texttt{scipy.optimize SLSQP} for both hardening and softening steps.
While the problem's solutions are not unique, the initial defect triggers localization with every test of the present implementation, although it was noted that the solutions were sensitive to the magnitude of the initial plastic strain.
In addition the strain step is almost systematically larger than parameter $\delta$, which is likely due to a potential snap--through mechanism in the proposed implementation.
This is illustrated in Figure~\ref{fig:localhard} where the work-hardening behavior of each finite element is plotted against the reference behavior.
There the solver tends to jump from the softening segment to the re-hardening segment when localization starts.
For the present class of models, this behavior can possibly be alleviated by using arc-length solvers or by altering the staggered algorithms which may tend to commit to the first picked minimizing direction.
Overall, the framework as described can reproduce plastic instabilities in a straightforward manner exploiting optimization strategies inherited from standard generalized materials properties.

This has the advantage of generality but requires some specialization to continue improving efficiency.
For example:
\begin{itemize}
 \item a conventional Newton method with a local return mapping algorithm for the hardening step will necessarily improve the algorithm's efficiency;
 \item the softening step can be improved by pre-computing softening variables solutions that disregard the lip constraint $\mathcal L$, and use them as initial guess or bound after performing a lip-projection \cite{chevaugeon2022}.
\end{itemize}
The observations are likely to hold when scaling this computational strategy to higher dimensions.


\paragraph{Model interpretability}

In the local potential \eqref{eqn:pizero}, the softening contribution can be expressed as 
\begin{align}
 \pi_s = \delta\sigma_a\left( (1-d)\bar p +\frac{d^{2}}{2}\right)
\end{align}
Integrating over a volume unaffected by plastic instabilities ($\bar p=d=0$) returns zero
while the integral over a volume $V$ with fully developed instability  ($\bar p=d=1$) gives 
\begin{align}
 \left.\Pi_s\right|_{d=1} = \int_V\left.\pi_{s}\right|_{d=1} \mathrm dv = V\frac{\delta\sigma_a}2.
\end{align}
This is the part of the plastic work corresponding to an instability. 
Parameters $\delta$ and $\sigma_a$ can be inferred from monotonic testing, and the energy above can be calibrated based on the regularization process and the profile of a localization event.
In the case of \myLF{} and a piecewise linear softening evolution (triangular localization profile of base $2\ell$, width $w$ and thickness $b$), then the plastic work of a localization band  is $\Pi_s=wb\ell(\delta\sigma_a)$.
The regularization length therefore maps to a measurable mechanical quantity.

The softening variable is field-agnostic but can be connected to the physics of different phenomena.
For example in strain it should link to densities of solutes during the (un)pinning of dislocations while in lattices or anticracks it will map to a change of density.
In general, $d$ will describe  a continuous transition from one phase to another, and there is a potential for each phenomenon to connect this macroscopic variable to structural or chemical processes.


\paragraph{Alternative regularization strategy}

The \myLF{} approach to softening has been compared to \myPF{} in the context of continuum damage \cite{gopalsamy2023}, and it was shown that both approaches yield comparable results while exhibiting distinct advantages.
For example \myLF{} regularization will not intervene in the problem resolution as long as softening is not triggered, while \myPF{} requires gradient computations over the entire domain at every time step.
On the other hand, \myPF{} requires minimal intervention at solver level (although gradient of internal variables are not handled by most commercial software out of the box) while \myLF{} demands dedicated implementations.
The present framework extends straightforwardly to \emph{gradient-enhanced} approaches by introducing a gradient contribution in the softening potential:
\begin{align}
\pi_s = \delta\sigma_a\left( (1-d)\bar p +\frac{d^{2}}{2} + \ell^2\|\nabla d\|^2\right)
\end{align}
where $\|\nabla d\|$ designates the norm of a gradient.
The internal length $\ell$ above holds similar properties in the rest of the present study and ultimately corresponds to the thickness of localization bands.
The present framework is therefore compatible with several popular regularization approaches inherited from the continuum damage community, while describing plastic instabilities.


\section{Conclusions}
This work demonstrates that plastic instabilities can be reformulated as a variational softening problem through the introduction of an explicit damage-like variable, thereby enabling direct use of regularization and solution strategies developed in ductile fracture mechanics.

To provide a minimal proof of concept, the proposed formulation uses linear elasticity, linear work-hardening, linear softening and a single localization event. 
However arbitrary hardening behavior can be used in the same potential, and it is shown in Appendix~\ref{sec:app} how changing the bar operator can extend the model to repeated softening (dynamic-strain-aging--like instabilities).

The broader ambition disclosed in introduction is a shared computational framework for softening problems across fields.
The natural next steps include tackling solver optimization and higher dimensionality, as well as field specific applications such as forming of sheet metals, and the compressive responses of architected and granular media.


%
%
%
%

\section*{Acknowledgements}

Prof.\ Nicolas Chevaugeon (\texttt{\textsc{orcid} 0000-0001-9123-1119}) is gratefully acknowledged for his help and valuable feedback on the method's implementation.

\appendix
\clearpage

\section{Analytical details}
\label{sec:app}

\subsection{State potential Hessian}
The potential
\begin{equation}
 \pi (
  \epsilon, \epsilon_p, p, d
 )
 = \frac12 E(\epsilon-\epsilon_p )^2
 + \sigma_y\left(p+\frac12kp^2\right)
 + \delta\left(1-d\right)\bar p\sigma_a
 + \delta\frac12d^2\sigma_a
\end{equation}
admits for Hessian:
\begin{align}
 \begin{pmatrix}
 \pderiv{}{\epsilon} \\ \pderiv{}{\epsilon_p} \\ \pderiv{}{p} \\ \pderiv{}{d}
 \end{pmatrix}
 \begin{pmatrix}
 \pderiv{}{\epsilon} \\ \pderiv{}{\epsilon_p} \\ \pderiv{}{p} \\ \pderiv{}{d}
 \end{pmatrix}^\top
 \pi 
 =
 \begin{bmatrix}
  E & -E & & \\
  -E & E & & \\
  &&k\sigma_y+\pderiv[2]{\bar p}{p}(1-d)\delta\sigma_a & -\pderiv{\bar p}{p}\delta\sigma_a \\
  && -\pderiv{\bar p}{p}\delta\sigma_a& \delta\sigma_a
 \end{bmatrix}
\end{align}

\subsection{Bar operator derivative}
The bar operator is used to connect the plastic evolution to the softening variable.
Its derivative is defined everywhere except in $p=\delta$. 
We use pragmatically
\begin{align}
 \bar p=\dfrac{\min{(p,\delta)}}{\delta},\qquad
 \pderiv{\bar p}{p} =\frac{1}{\delta}\,\mathcal I_{(p<\delta)}(p),\qquad
 \pderiv[2]{\bar p}{p} =0;
\end{align}
with $\mathcal I$ the indicator function.

\subsection{Alternative bar operator definition for recurring softening}
Let 
\begin{align}
 \left.
  \begin{array}{rll} 
  \bar p:\mathbb R^+&\longrightarrow& [0,1]\\
  p&\longmapsto&\dfrac{p\,\mathrm{mod}\, \delta}{\delta}
 \end{array}
 \right]
\end{align}
Its derivative is defined almost everywhere:
\begin{align}
 \pderiv{\bar p}{p} =\frac{1}{\delta},\qquad
 \pderiv[2]{\bar p}{p} =0,\qquad
 \quad\forall p\neq k\delta, \quad k\in\mathbb N.
\end{align}

\bibliographystyle{elsarticle-harv} 

\begin{thebibliography}{17}
\expandafter\ifx\csname natexlab\endcsname\relax\def\natexlab#1{#1}\fi
\providecommand{\url}[1]{\texttt{#1}}
\providecommand{\href}[2]{#2}
\providecommand{\path}[1]{#1}
\providecommand{\DOIprefix}{doi:}
\providecommand{\ArXivprefix}{arXiv:}
\providecommand{\URLprefix}{URL: }
\providecommand{\Pubmedprefix}{pmid:}
\providecommand{\doi}[1]{\href{http://dx.doi.org/#1}{\path{#1}}}
\providecommand{\Pubmed}[1]{\href{pmid:#1}{\path{#1}}}
\providecommand{\bibinfo}[2]{#2}
\ifx\xfnm\relax \def\xfnm[#1]{\unskip,\space#1}\fi
\bibitem[{Barraclough et~al.(2016)Barraclough, Blackford, Liebenstein,
  Sandfeld, Stratford, Weinl{\"{a}}nder and Zaiser}]{barraclough2016}
\bibinfo{author}{Barraclough, T.W.}, \bibinfo{author}{Blackford, J.R.},
  \bibinfo{author}{Liebenstein, S.}, \bibinfo{author}{Sandfeld, S.},
  \bibinfo{author}{Stratford, T.J.}, \bibinfo{author}{Weinl{\"{a}}nder, G.},
  \bibinfo{author}{Zaiser, M.}, \bibinfo{year}{2016}.
\newblock \bibinfo{title}{Propagating compaction bands in confined compression
  of snow}.
\newblock \bibinfo{journal}{Nature Physics} \bibinfo{volume}{13},
  \bibinfo{pages}{272--275}.
\newblock \URLprefix \url{https://arxiv.org/abs/1501.02184},
  \DOIprefix\doi{10.1038/nphys3966}.
\bibitem[{Bertoldi and Boyce(2008)}]{bertoldi2008}
\bibinfo{author}{Bertoldi, K.}, \bibinfo{author}{Boyce, M.C.},
  \bibinfo{year}{2008}.
\newblock \bibinfo{title}{Wave propagation and instabilities in monolithic and
  periodically structured elastomeric materials undergoing large deformations}.
\newblock \bibinfo{journal}{Physical Review B} \bibinfo{volume}{78},
  \bibinfo{pages}{184107}.
\newblock \DOIprefix\doi{10.1103/physrevb.78.184107}.
\bibitem[{Bourdin et~al.(2008)Bourdin, Francfort and Marigo}]{bourdin2008}
\bibinfo{author}{Bourdin, B.}, \bibinfo{author}{Francfort, G.A.},
  \bibinfo{author}{Marigo, J.J.}, \bibinfo{year}{2008}.
\newblock \bibinfo{title}{The variational approach to fracture}.
\newblock \bibinfo{journal}{Journal of Elasticity} \bibinfo{volume}{91},
  \bibinfo{pages}{5--148}.
\newblock \DOIprefix\doi{10.1007/s10659-007-9107-3}.
\bibitem[{Chevaugeon and Moës(2022)}]{chevaugeon2022}
\bibinfo{author}{Chevaugeon, N.}, \bibinfo{author}{Moës, N.},
  \bibinfo{year}{2022}.
\newblock \bibinfo{title}{Lipschitz regularization for fracture: The lip-field
  approach}.
\newblock \bibinfo{journal}{Computer Methods in Applied Mechanics and
  Engineering} \bibinfo{volume}{402}, \bibinfo{pages}{115644}.
\newblock \URLprefix \url{http://dx.doi.org/10.1016/j.cma.2022.115644},
  \DOIprefix\doi{10.1016/j.cma.2022.115644}.
\bibitem[{Gaume et~al.(2018)Gaume, Gast, Teran, van Herwijnen and
  Jiang}]{gaume2018}
\bibinfo{author}{Gaume, J.}, \bibinfo{author}{Gast, T.},
  \bibinfo{author}{Teran, J.}, \bibinfo{author}{van Herwijnen, A.},
  \bibinfo{author}{Jiang, C.}, \bibinfo{year}{2018}.
\newblock \bibinfo{title}{Dynamic anticrack propagation in snow}.
\newblock \bibinfo{journal}{Nature Communications} \bibinfo{volume}{9}.
\newblock \DOIprefix\doi{10.1038/s41467-018-05181-w}.
\bibitem[{Gopalsamy et~al.(2023)Gopalsamy, Chevaugeon, Chupin and
  Hammoum}]{gopalsamy2023}
\bibinfo{author}{Gopalsamy, R.}, \bibinfo{author}{Chevaugeon, N.},
  \bibinfo{author}{Chupin, O.}, \bibinfo{author}{Hammoum, F.},
  \bibinfo{year}{2023}.
\newblock \bibinfo{title}{Variational approach to viscoelastic fracture:
  comparison of a phase-field and a lip-field approach}.
\newblock \bibinfo{journal}{International Journal of Fracture}
  \bibinfo{volume}{244}, \bibinfo{pages}{163--185}.
\newblock \DOIprefix\doi{10.1007/s10704-023-00725-8}.
\bibitem[{Halphen and Son~Nguyen(1975)}]{halphen1975}
\bibinfo{author}{Halphen, B.}, \bibinfo{author}{Son~Nguyen, Q.},
  \bibinfo{year}{1975}.
\newblock \bibinfo{title}{{Sur les mat{\'e}riaux standard
  g{\'e}n{\'e}ralis{\'e}s}}.
\newblock \bibinfo{journal}{{Journal de M{\'e}canique}} \bibinfo{volume}{14},
  \bibinfo{pages}{39--63}.
\newblock \URLprefix \url{https://hal.science/hal-03600755}.
\bibitem[{Kubin and Estrin(1985)}]{kubin1985}
\bibinfo{author}{Kubin, L.P.}, \bibinfo{author}{Estrin, Y.},
  \bibinfo{year}{1985}.
\newblock \bibinfo{title}{The {Portevin-Le Chatelier} effect in deformation
  with constant stress rate}.
\newblock \bibinfo{journal}{Acta Metallurgica} \bibinfo{volume}{33},
  \bibinfo{pages}{397--407}.
\newblock \DOIprefix\doi{10.1016/0001-6160(85)90082-3}.
\bibitem[{Manach et~al.(2014)Manach, Thuillier, Yoon, Co{\"e}r and
  Laurent}]{manach2014}
\bibinfo{author}{Manach, P.Y.}, \bibinfo{author}{Thuillier, S.},
  \bibinfo{author}{Yoon, J.W.}, \bibinfo{author}{Co{\"e}r, J.},
  \bibinfo{author}{Laurent, H.}, \bibinfo{year}{2014}.
\newblock \bibinfo{title}{Kinematics of {Portevin-Le Chatelier} bands in simple
  shear}.
\newblock \bibinfo{journal}{International Journal of Plasticity}
  \bibinfo{volume}{58}, \bibinfo{pages}{66--83}.
\newblock \DOIprefix\doi{10.1016/j.ijplas.2014.02.005}.
\bibitem[{Marais et~al.(2012)Marais, Mazi{\`e}re, Forest, Parrot and
  Le~Delliou}]{marais2012}
\bibinfo{author}{Marais, A.}, \bibinfo{author}{Mazi{\`e}re, M.},
  \bibinfo{author}{Forest, S.}, \bibinfo{author}{Parrot, A.},
  \bibinfo{author}{Le~Delliou, P.}, \bibinfo{year}{2012}.
\newblock \bibinfo{title}{Identification of a strain-aging model accounting for
  {L\"uders} behavior in a {C-Mn} steel}.
\newblock \bibinfo{journal}{Philosophical magazine} \bibinfo{volume}{92},
  \bibinfo{pages}{3589--3617}.
\newblock \DOIprefix\doi{10.1080/14786435.2012.699687}.
\bibitem[{McCormick(1988)}]{mccormick1988}
\bibinfo{author}{McCormick, P.G.}, \bibinfo{year}{1988}.
\newblock \bibinfo{title}{Theory of flow localisation due to dynamic strain
  ageing}.
\newblock \bibinfo{journal}{Acta Metallurgica} \bibinfo{volume}{36},
  \bibinfo{pages}{3061--3067}.
\newblock \DOIprefix\doi{10.1016/0001-6160(88)90043-0}.
\bibitem[{Meijer et~al.(2026)Meijer, Shaik, McDermott and Jaeger}]{meijer2026}
\bibinfo{author}{Meijer, J.G.}, \bibinfo{author}{Shaik, F.H.},
  \bibinfo{author}{McDermott, V.V.}, \bibinfo{author}{Jaeger, H.M.},
  \bibinfo{year}{2026}.
\newblock \bibinfo{title}{Diffusive buckling fronts in lattice-based
  metamaterials}.
\newblock \bibinfo{journal}{PNAS Nexus} \bibinfo{volume}{5}.
\newblock \DOIprefix\doi{10.1093/pnasnexus/pgag272}.
\bibitem[{Miehe et~al.(2010)Miehe, Welschinger and Hofacker}]{miehe2010}
\bibinfo{author}{Miehe, C.}, \bibinfo{author}{Welschinger, F.},
  \bibinfo{author}{Hofacker, M.}, \bibinfo{year}{2010}.
\newblock \bibinfo{title}{Thermodynamically consistent phase‐field models of
  fracture: Variational principles and multi‐field fe implementations}.
\newblock \bibinfo{journal}{International Journal for Numerical Methods in
  Engineering} \bibinfo{volume}{83}, \bibinfo{pages}{1273--1311}.
\newblock \DOIprefix\doi{10.1002/nme.2861}.
\bibitem[{Moës and Chevaugeon(2021)}]{moes2021}
\bibinfo{author}{Moës, N.}, \bibinfo{author}{Chevaugeon, N.},
  \bibinfo{year}{2021}.
\newblock \bibinfo{title}{Lipschitz regularization for softening material
  models: the {Lip}-field approach}.
\newblock \bibinfo{journal}{Comptes Rendus M\'{e}canique}
  \bibinfo{volume}{349}, \bibinfo{pages}{415--434}.
\newblock \DOIprefix\doi{10.5802/crmeca.91}.
\bibitem[{Reyne(2026)}]{gitlabplasmat}
\bibinfo{author}{Reyne, B.}, \bibinfo{year}{2026}.
\newblock \bibinfo{title}{Software implementation}.
\newblock \bibinfo{howpublished}{GitLab}.
\newblock \URLprefix \url{gitlab.sintef.no/plastic{\_}instabilities/plasmat}.
\bibitem[{Rowlands et~al.(2023)Rowlands, Rae and Galindo-Nava}]{rowlands2023}
\bibinfo{author}{Rowlands, B.S.}, \bibinfo{author}{Rae, C.},
  \bibinfo{author}{Galindo-Nava, E.}, \bibinfo{year}{2023}.
\newblock \bibinfo{title}{The portevin-le chatelier effect in nickel-base
  superalloys: Origins, consequences and comparison to strain ageing in other
  alloy systems}.
\newblock \bibinfo{journal}{Progress in Materials Science}
  \bibinfo{volume}{132}, \bibinfo{pages}{101038}.
\newblock \DOIprefix\doi{10.1016/j.pmatsci.2022.101038}.
\bibitem[{Tancogne-Dejean et~al.(2016)Tancogne-Dejean, Spierings and
  Mohr}]{tancognedejean2016}
\bibinfo{author}{Tancogne-Dejean, T.}, \bibinfo{author}{Spierings, A.B.},
  \bibinfo{author}{Mohr, D.}, \bibinfo{year}{2016}.
\newblock \bibinfo{title}{Additively-manufactured metallic micro-lattice
  materials for high specific energy absorption under static and dynamic
  loading}.
\newblock \bibinfo{journal}{Acta Materialia} \bibinfo{volume}{116},
  \bibinfo{pages}{14--28}.
\newblock \DOIprefix\doi{10.1016/j.actamat.2016.05.054}.

\end{thebibliography}





\end{document}